# Long-lived and multiplexed atom-photon entanglement interface with feed-forward-controlled readouts

Shengzhi Wang, Minjie Wang, Yafei Wen, Zhongxiao Xu, Tengfei Ma,

Shujing Li, Hai Wang*

*The State Key Laboratory of Quantum Optics and Quantum Optics Devices, Institute of Opto-Electronics, Shanxi University, Taiyuan 030006, China*

*Collaborative Innovation Center of Extreme Optics, Shanxi University, Taiyuan 030006, China*

The quantum interface (QI) that generates entanglement between photonic and spin-wave (atomic memory) qubits is a basic building block for quantum repeaters. Realizing ensemble-based repeaters in practice requires quantum memory providing long lifetime and multimode capacity. Significant progresses have been achieved on these separate goals. The remaining challenge is to combine long-lived and multimode memories into a single QI. Here, by establishing multimode, magnetic-field-insensitive and long-wavelength spin-wave storage in laser-cooled atoms that are placed inside a phase-passively-stabilized polarization interferometer, we constructed a multiplexed QI that stores up to three long-lived spin-wave qubits. Using a feed-forward-controlled system, we demonstrated that the multiplexed QI gives rise to a 3-fold increase in the atom-photon (photon-photon) entanglement-generation probability compared to single-mode QIs. The measured Bell parameter is $2.5 \pm 0.1$ combined with a memory lifetime up to 1ms. The presented work represents a key step forward in realizing fiber-based long-distance quantum communications.

Quantum repeaters (QRs) [1] hold promise for distributing entanglement over long distances ( >1000 $km$) via optical fibers, thereby providing a feasible path to realize long-distance quantum communications [2-4] and quantum networks [5, 6]. In QRs, long distances are divided into short elementary links, with each link comprising two nodes that store quantum states [1-3]. For each link, entanglement between two nodes is required to be established in a "heralded" way [2, 4]. Various physical systems, such as atomic ensembles [2-4, 7, 8] and single quantum systems, including single atoms [9-10], ions [11-12] and solid-state spins [13-14], have been proposed as the nodes. The atomic-ensemble-based nodes, which were initially proposed in the Duan-Lukin-Cirac-Zoller (DLCZ) protocol [7], are formed by QIs that create quantum correlations between a spin wave (SW) stored in the atomic ensembles and a photon via spontaneous Raman emissions (SREs) [2, 15-28]. The quantum correlations created via SREs form the basis of generating entanglement between a photonic qubit and a spin-wave qubit [2, 29-32]. Ensemble-based QIs are attractive because large number of atoms ensures an efficient quantum memory (QM) [2, 18-19]. In an improved DLCZ scheme [33-34], the QR uses the spin-wave-photon entanglement (SWPE) instead of quantum correlations as nodes, thereby removing the requirement for long-distance phase stability [35] in the original DLCZ protocol. Over the past decade, QIs that generate spin-wave-photon (atom-photon) entanglement through SREs [29-32, 36–41] or storage of photonic entanglement [42–44] in atomic ensembles have been demonstrated. With atom-photon entanglement, quantum teleportation [45–47] and entanglement generations [35, 48] in elementary links have been demonstrated.

In QRs, QMs are required to have long lifetimes to store the generated entanglement in elementary links [2-4, 49]. To achieve long-lived DLCZ-like QMs, the decoherence of the SWs in cold atoms was widely studied [20-26, 37]. Atomic motions and inhomogeneous broadening of the spin transitions were shown to causes spin-wave dephasing. Motion-induced decoherence was suppressed either using collinear configuration to lengthen spin-wave wavelengths [22-23] or confining the atoms in optical lattices [24-26, 37]. Inhomogeneous-broadening-induced decoherence may be reduced using magnetic-field-insensitive (MFI) coherences for spin-wave storage [22-26, 37]. Long-lived (0.1-$s$) and non-multiplexed atom-photon entanglement QI was demonstrated in optical-lattice atoms [37], in which the memory qubit was stored as two spatially-distinct SWs, both associated with the $0 \leftrightarrow 0$ MFI (clock) coherence, and the corresponding photonic qubit encoded into two arms of a Mach-Zehnder interferometer. However, to maintain maximal entanglement in the experiment, the relative phase between the arms was actively stabilized to zero by coupling an auxiliary laser beam into the interferometer [37]. In consequence, the experimental setup was technically complex and difficult to scale.

DLCZ-type QRs using single-mode elementary links have been recognized to have very slow rates for practical use [2, 50-54]. To overcome this problem, one may use multimode QMs instead of single-mode ones in the QIs (nodes) to increase entanglement-generation rate in elementary links [2, 4, 26, 48, 52-56]. Multimode quantum storages of single SWs have been mainly implemented with rare-earth-ion-doped (REID) crystals [57-59] or cold atoms [56, 60-65]. In cold atoms, spatial [56, 63-65] and temporal [66] multimode entanglement between photonic and atomic memory qubits have been generated via spatial mode operations. Based on temporal multimode DLCZ-like quantum memories in REID crystals [57, 58], K. Kutluer *et. al.* experimentally demonstrated time-bin entanglement between a SW and a photon [32]. With more modes being used in the experiment, multimode entanglement in time will be generated with the crystal. Additionally, continuous-variable entanglement between light and a crystal has been generated in two temporal modes [67]. However, to date, the durations for preserving multimode entanglement are below 50 $\mu s$. Limited to this lifetime, the entanglement creation between two multimode QMs linked by more than 10-*km* fibers can't be established in the "heralded" way (see Supplementary Material [68]). The realizations of QRs using multimode SWPE QIs require QMs to have long storage durations and multiplexed qubit storages [3, 4, 48, 56], more specifically, to store each of the multiplexed atomic qubits individually as a superposition of long-lived spin waves. However, that goal remained elusive.

In the present experiment, we overcome the difficulties by integrating multimode, MFI and long-wavelength spin-wave storage in a single QM system. The QM system was formed from an ensemble of laser-cooled $^{87}$Rb atoms placed in a polarization interferometer (PI). The PI was formed using two identical beam displacers (BDs). Three optical channels (OCs) across the PI were built for multimode storages. The two arms of each OC, which correspond to the *H*- and *V*- polarization modes of the BDs, were used to encode photonic qubits. The relative phase between the paired arms was passively stable [69-72]. The six ($3\times2$) spatial modes arranged in a two-dimensional array were focused at the center of the atoms with a lens. The atomic excitations, created by SREs, are stored as MFI spin waves of long wavelengths. We then realized an MQI that generated long-lived SWEP in the three channels.

The cold atomic ensemble was centered in a PI formed by BD1 and BD2 (see Fig.1a). The experiment relied on SREs induced by write pulses propagating along *z*-axis to create entangled pairs, each pair comprising a Stokes photon and a spin-wave excitation (cf. below). To realize the MQI, we set up three optical channels (spatial modes) that go through the PI to collect and detect both the Stokes and retrieved photons. The three channels (labeled by $OC_{i=1,2,3}$) are arranged in a vertical plane with a separation of 4 *mm*. Each channel is pre-aligned with light beam. For example, the light beam in $OC_i$ emitted from the *i*-th single-mode fiber at left site (labeled by $SMF_T^{(i)}$), enters BD1, which split the *H*- and *V*-polarization components of the beam into two modes $A_1^{(i)}$ and $A_2^{(i)}$, corresponding to the two arms of $OC_i$. Exiting from BD1, the two arms parallel propagate in a horizontal plane, with the same separation of 4 mm. Hence, there are six spatial modes ( $A_\alpha^{(i)}$ with $\alpha$=1, 2; $i$ = 1 to 3) which are arranged parallel in a two-dimensional array (see *L*-section of the array in Fig.1b). The optical elements (Fig.1a) including the two identical lenses (L1 and L2) and two beam transformation devices (BTD1 and BTD2), are inserted in the PI (see Supplementary Material [68]), here, BTD1 (BTD2) is formed by two lenses, that shrink (expand) the beam array by a factor $F$ (see Supplementary Material [68]). The effective multimode storages rely on strong couplings of the Stokes and retrieved photons with the atoms. To this end, we use lens L1 to focus the six modes at the center of the atoms. To ensure the multimode storages have long lifetimes, we have to store long-wavelength spin waves, which in turn require the angles $\vartheta_{A_\alpha}^{(i)}$ of the six modes $A_\alpha^{(i)}$ relative to the write beam to be reduce very small values [21, 73]. The angles are calculated from $\vartheta_{A_\alpha}^{(i)} = \left(\sqrt{4(i-2)^2 + 1}\right) B_f / 2f$ (see Supplementary Material [68]), where $B_f$ denotes the beam separation of the array on the lens L1, and $f$ the focal length of L1. To reduce the values of angles significantly, we selected $f = 1.425\ m$ and used BTD1 to reduce the array beam separation by factor $F=2$. After BTD1, the array propagates parallel to L1 and has a separation of $B_f = 2\ mm$. We then obtain small angles $\{\vartheta_{A_1}^{(1)} = \vartheta_{A_2}^{(1)} \approx 0.09°, \vartheta_{A_1}^{(2)} = \vartheta_{A_2}^{(2)} \approx 0.04°,\ \vartheta_{A_1}^{(3)} = \vartheta_{A_2}^{(3)} \approx 0.09°\}$, which correspond to lifetimes limited by the atomic motion of $\{840\ \mu s,\ 1850\ \mu s,\ 840\ \mu s\}$ for modes $\{A_{1,2}^{(1)},\ A_{1,2}^{(2)},\ A_{1,2}^{(3)}\}$, respectively (see Supplementary Material [68]). Additionally, the spot size of the array at the atomic center is 0.65 *mm*, which is much less than the atomic transverse size (2 *mm*). After passing through the atoms, the six crossed beams are transformed to a parallel beam array by L2. Then, the array goes through BTD2 and is expanded by factor *F=2*. After the transformation, this array has the same beam separations and sizes as that depicted in Fig.1b (see Supplementary Material [68]). Next, the array passes through BD2, which combines the paired arm modes into single spatial modes; for example, $A_1^{(i)}$ and $A_2^{(i)}$ modes are combined into single light beam in OC$_i$. Finally, the light beam in OC$_i$ is coupled into the *i*-th single-mode fiber at right site (labeled SMF$_S^{(i)}$ in Fig. 1a) with high efficiencies (see Supplementary Material [68]).

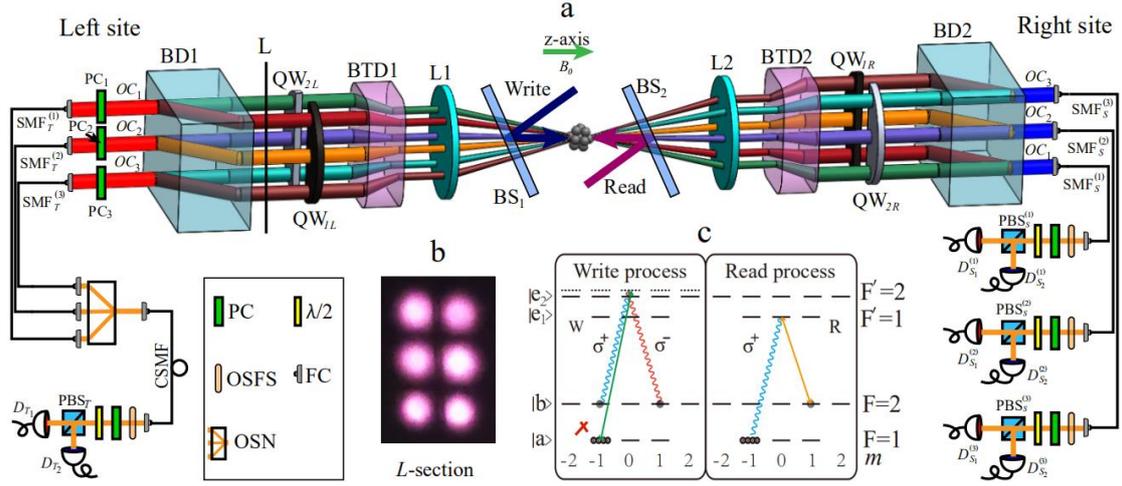

Fig.1 Overview of the experiment. (a) Experiment setup for the 3-mode MQI. PC: phase compensator (see Supplementary Material [68]); OSN: optical switching network; QW: λ/4 wave-plate; BD: beam displacer, PBS: polarization beam splitter; BTD: beam transformation device; SMF: single-mode fiber; OC: optical channel. BS1 (BS2): Non-polarizing beam splitter, whose reflectance (transmission) is 10% (90%). The write beam is coupled into the z-axis via BS1, and the read beam into the opposite direction to that of the write beam via BS2; OSFS: optical-spectrum-filter set (see Supplementary Material [68]). $B_0$: bias magnetic field (4G). (b) Relevant atomic levels. (c) Pattern of the array at L-section.

The relevant Rb atomic levels (Fig. 1c) are $|a\rangle = |5^2S_{1/2}, F=1\rangle$, $|b\rangle = |5^2S_{1/2}, F=2\rangle$, $|e_1\rangle = |5^2P_{1/2}, F'=1\rangle$ and $|e_2\rangle = |5^2P_{1/2}, F'=2\rangle$. After the atoms are prepared in the Zeeman state $|a, m_a=-1\rangle$ via optical pumping [74], we start the SWPE generation. At the beginning of a trail, a writing pulse with 20MHz blue-detuned to the $|a\rangle \to |e_2\rangle$ transition is applied to the atoms. The write pulse induces the Raman transition $|a, m_a = -1\rangle \rightarrow |b, m_b = 1\rangle$ ($|a, m_a = -1\rangle \rightarrow |b, m_b = -1\rangle$) via $|e_2, m' = 0\rangle$, which may emit $\sigma^-$-polarized ($\sigma^+$-polarized) Stokes photons and create simultaneously spin-wave excitations associated with the coherence $|m_a = -1\rangle \leftrightarrow |m_b = 1\rangle$ ($|m_a = -1\rangle \leftrightarrow |m_b = -1\rangle$) (Fig.1c), where $|m_a = -1\rangle \leftrightarrow |m_b = 1\rangle$ and $|m_a = -1\rangle \leftrightarrow |m_b = -1\rangle$ are the MFI and the magnetic-field-sensitive (MFS) coherences, respectively. If the Stokes photon emits into the $A_1^{(i)}$ ($A_2^{(i)}$) mode and moves rightwards, it is denoted as $S_1^{(i)}$ ($S_2^{(i)}$). In this case, one excitation is created in the spin-wave mode $M_1^{(i)}$ ($M_2^{(i)}$) defined by the wave-vector $k_{M_1}^{(i)} = k_W - k_{S_1}^{(i)}$ ($k_{M_2}^{(i)} = k_W - k_{S_2}^{(i)}$), where $k_W$ denotes the wave-vector of the write pulse, and $k_{S_1}^{(i)}$ ($k_{S_2}^{(i)}$) that of the Stokes photon $S_1^{(i)}$ ($S_2^{(i)}$). In the Fig. 1a, $\sigma^-$-polarized $S_1^{(i)}$ ($S_2^{(i)}$) photons are transformed into $H$ ($V$) –polarized photons by the λ/4 plate labeled as QW$_{1R}$ (QW$_{2R}$). After BD2, the $H$ ($V$)-polarized $S_1^{(i)}$ and $S_2^{(i)}$ modes are combined to form a Stokes qubit $S^{(i)}$ and then are coupled to SMF$_S^{(i)}$. In addition, the corresponding excitations in the $M_1^{(i)}$ and $M_2^{(i)}$ modes are stored as MFI spin waves, which represent the $i$-th atomic qubit. In the present experiment, the excitation probability $\chi$ for each $OC$ is almost the same, i.e., $\chi_1 \approx \chi_2 \approx \chi_3 \approx \chi$. For $\chi \ll 1$, the entangled state between the $i$-th atomic and photonic qubits is described as $\Phi_{\text{a-p}}^{i\text{-th}} = |H\rangle_S^{(i)}|M_1\rangle_{FI}^{(i)} + e^{i\varphi_i}|V\rangle_S^{(i)}|M_2\rangle_{FI}^{(i)}$, where $|H\rangle_S^{(i)}$ ($|V\rangle_S^{(i)}$) denotes the $H$ ($V$) -polarized Stokes photon of the qubit $S^{(i)}$, $|M_1\rangle_{FI}^{(i)}$ ($|M_2\rangle_{FI}^{(i)}$) one MFI excitation in the spin-wave modes $M_1^{(i)}$ ($M_2^{(i)}$), and $\varphi_i$ the phase difference between the $S_1^{(i)}$ and $S_2^{(i)}$ fields. If the $S_1^{(i)}$ ($S_2^{(i)}$) photon is $\sigma^+$-polarized, the corresponding excitation in the $M_1^{(i)}$ ($M_2^{(i)}$) mode is stored as the MFS spin wave and decays rapidly [75]. However, these photons are abandoned because they are excluded from the collections (see Supplementary Material [68]).

Returning to the entangled state $\Phi_{\text{a-p}}^{i\text{-th}}$, the qubit $S^{(i)}$ is guided into the $i$-th polarization-beam splitter ($\text{PBS}_S^{(i)}$) after the $\text{SMF}_S^{(i)}$. The two outputs of the $\text{PBS}_S^{(i)}$ are sent to single-photon detectors $D_{S_1}^{(i)}$ and $D_{S_2}^{(i)}$. The polarization angle of qubit $S^{(i)}$, denoted by $\theta_{S_i}$, may be changed by rotating the $\lambda/2$-plate before the $\text{PBS}_S^{(i)}$. Here, we set $\theta_{S_1} = \theta_{S_2} = \theta_{S3} = \theta_S$. Once a photon is detected by $D_{S_1}^{(i)}$ ($D_{S_2}^{(i)}$), the storage of a spin-wave excitation $|M_1\rangle_{FI}^{(i)}$ ($|M_2\rangle_{FI}^{(i)}$) is heralded. After a storage time $t$, we apply a reading pulse that counter-propagates with the write beam to convert the spin-wave excitation $|M_1\rangle_{FI}^{(i)}$ ($|M_2\rangle_{FI}^{(i)}$) into an anti-Stokes photon $T_1^{(i)}$ ($T_2^{(i)}$). The retrieved photon $T_1^{(i)}$ ($T_2^{(i)}$) is emitted into the spatial mode determined by the wave-vector constraint $k_{T_1}^{(i)} \approx -k_{S_1}^{(i)}$ ($k_{T_2}^{(i)} \approx -k_{S_2}^{(i)}$); i.e., it propagates in arm $A_1^{(i)}$ ($A_2^{(i)}$) along the opposite direction to the $S_1^{(i)}$ ($S_2^{(i)}$) photon. The $T_1^{(i)}$ ($T_2^{(i)}$) photon is $\sigma^+$-polarized and transformed into the $H$ ($V$)-polarized photon by a $\lambda/4$ plate labeled $\text{QW}_{1L}$ ($\text{QW}_{2L}$). After BD1, the $T_1^{(i)}$ and $T_2^{(i)}$ fields are combined to form a polarization qubit ($T^{(i)}$). Thus, the atom–photon state $\Phi_{a-p}^{(i)}$ is transformed into the two-photon entangled state, $\Phi_{\text{pp}}^{i\text{-th}} = \left|H_S^{(i)}\right\rangle\left|H_T^{(i)}\right\rangle + e^{i(\varphi_i + \psi_i)}\left|V_S^{(i)}\right\rangle\left|V_T^{(i)}\right\rangle$, where $\psi_i$

denotes the phase difference between the anti-Stokes fields in arms $A_1^{(i)}$ and $A_2^{(i)}$ before they overlap at BD1. Using the *i*-th phase compensator (labeled PC$_i$ in Fig. 1a), we set the phase difference $\varphi_i + \psi_i$ to zero. The generation of atom–photon (photon–photon) entanglement based on *m*=3 storage modes constitutes the MQI operation. To enable the MQI to be available for the multiplexed QR scheme [56], we introduced an optical switch network (OSN) to route the retrieved qubits into a common single-mode-fiber [56] (CSMF). Passing through the CSMF and a $\lambda/2$ plate, the qubits $T^{(i)}$ ($i$ = 1 to 3) impinge on a polarization-beam splitter, PBS$_T$. The two outputs of the PBS$_T$ are sent separately to detectors $D_{T1}$ and $D_{T2}$. Then, the atoms are prepared in the initial state via optical pumping [74]. If no Stokes photon is detected during the write pulse, the atoms are pumped directly back into the initial state. Subsequently, the next trial starts.

To show that the MQI provides long-lived spin-wave storage, we examined the dependence of the retrieval efficiency on storage time *t*. The retrieval efficiency of the *m*-mode MQI is measured as $\gamma^{(m)} = \sum_{i=1}^{m} P_{S,T}^{i\text{-th}} / \left(\eta_T \sum_{i=1}^{m} P_S^{i\text{-th}}\right)$, where, where $\eta_T$ denotes the detection efficiency in the anti-Stokes channel, $P_{S,T}^{i\text{-th}} = P_{D_{S_1},D_{T_1}}^{i\text{-th}} + P_{D_{S_2},D_{T_2}}^{i\text{-th}}$ the Stokes-anti-Stokes coincidence probability, $P_{D_{S_1},D_{T_1}}^{i\text{-th}}$ ($P_{D_{S_2},D_{T_2}}^{i\text{-th}}$) the probability of detecting a coincidence between the detectors $D_{S_1}^{(i)}$ ($D_{S_2}^{(i)}$) and $D_{T_1}$ ($D_{T_2}$), $P_S^{i\text{-th}} = P_{D_{S_1}}^{i\text{-th}} + P_{D_{S_2}}^{i\text{-th}}$ the Stokes-detection probability, $P_{D_{S_1}}^{i\text{-th}}$ ($P_{D_{S_2}}^{i\text{-th}}$) the probability of detecting a photon at $D_{S_1}^{(i)}$ ($D_{S_2}^{(i)}$), both $P_{S,T}^{i\text{-th}}$ and $P_S^{i\text{-th}}$ are measured for $\theta_S = \theta_T = 0^0$, and $\theta_T$ is the polarization angle of the $T^{(i)}$ qubits. The measured results for the MQI (black dots in Fig. 2) are based on storages of the three spin-wave qubits. The fitting function (solid red curve) based on $\gamma^{(m=3)}(t) = \gamma_0 e^{-t/\tau_0}$ yields a zero-delay retrieval efficiency $\gamma_0 \approx 15\%$ and $1/e$ storage time $\tau_0 \approx 870\ \mu s$. This lifetime is in agreement with the average lifetime over the three spin-wave qubits (see Supplementary Material [68]).

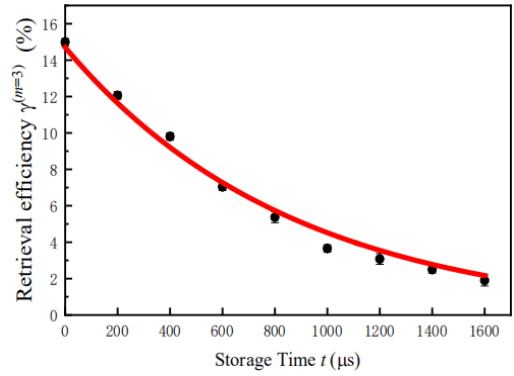

Fig.2 Retrieval efficiency of the three-mode MQI as a function of storage time $t$ for $\chi = 1\%$.

The quality of the $m$-mode SWPE is described by the Clauser-Horne-Shimony-Holt (CHSH) Bell parameter $S^{(m)}$ [56] written as:

$S^{(m)} = |E^{(m)}(\theta_S, \theta_T) - E^{(m)}(\theta_S, \theta_T') + E^{(m)}(\theta_S', \theta_T) + E^{(m)}(\theta_S', \theta_T')| < 2$ with the correlation function $E^{(m)}(\theta_S, \theta_T)$ defined by:

$$\frac{\sum_{i=1}^{m}\left[C_{D_{S_1},D_{T_1}}^{i\text{-th}}(\theta_S,\theta_T)+C_{D_{S_2},D_{T_2}}^{i\text{-th}}(\theta_S,\theta_T)-C_{D_{S_1},D_{T_2}}^{i\text{-th}}(\theta_S,\theta_T)-C_{D_{S_2},D_{T_1}}^{i\text{-th}}(\theta_S,\theta_T)\right]}{\sum_{i=1}^{m}\left[C_{D_{S_1},D_{T_1}}^{i\text{-th}}(\theta_S,\theta_T)+C_{D_{S_2},D_{T_2}}^{i\text{-th}}(\theta_S,\theta_T)+C_{D_{S_1},D_{T_2}}^{i\text{-th}}(\theta_S,\theta_T)+C_{D_{S_2},D_{T_1}}^{i\text{-th}}(\theta_S,\theta_T)\right]}, \qquad (1)$$

where, for example, $C_{D_{S_1},D_{T_1}}^{i\text{-th}}(\theta_S,\theta_T)$ ($C_{D_{S_2},D_{T_2}}^{i\text{-th}}(\theta_S,\theta_T)$) is coincidence counts between the detectors $D_{S_1}^{(i)}$ ($D_{S_2}^{(i)}$) and $D_{T_1}$ ($D_{T_2}$) for the polarization angles $\theta_S$ and $\theta_T$. In the $S^{(m)}$ measurement, we use the canonical settings $\theta_S = 0^0$, $\theta'_S = 45^0$, $\theta_T = 22.5^0$, $\theta'_T = 67.5^0$. To demonstrate that our three-mode MQI preserves entanglement over a long duration, we measured the decay of the parameter $S^{(m=3)}$ for various storage times $t$ (blue squares in Fig. 3). At $t = 1\ ms$, $S^{(m=3)} = 2.07 \pm 0.02$, which violates the CHSH inequality by 3.5 standard deviations.

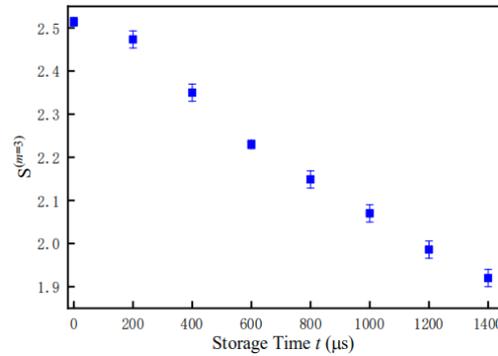

Fig.3 Bell parameters $S^{(m)}$ as a function of $t$ for $\chi = 1\%$. Error bars represent 1 standard deviation.

The quality of the photon-photon (atom-photon) entanglement generated from the $m$-mode MQI can also be characterized by the fidelity, which is given by $F^{(m)} = \mathrm{Tr}\left(\sqrt{\sqrt{\rho_r^{(m)}}\rho_d\sqrt{\rho_r^{(m)}}}\right)^2$, where, $\rho_r^{(m)}$ ($\rho_d$) denotes the reconstructed (ideal) density matrix of the photon–photon entangled state. Based on measurements of Stokes–anti-Stokes coincidences for $t = 1\ \mu s$ and $\chi = 1\%$, we reconstructed $\rho_r^{(m=3)}$ (see Supplementary Material [68]), which yields $F^{(m=3)} = 90.4 \pm 1.6\%$. We also reconstructed the density matrices $\rho_r^{i\text{-th}}$ of the entangled states $\Phi_{pp}^{i\text{-th}}$, which yields fidelities $F^{(1\text{-th})} = 88.6 \pm$

1.13% , $F^{(2\text{-th})} = 92.0 \pm 1.5\%$ and $F^{(3\text{-th})} = 88.4 \pm 0.85\%$ (see Supplementary Material [68]). Their average fidelity $\bar{F} = 89.7 \pm 1.2\%$ is in agreement with the value of $F^{(m=3)}$.

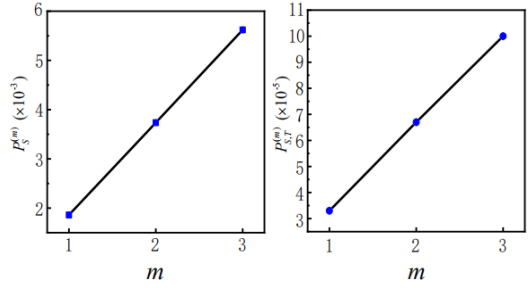

Fig.4 Stokes detection probability (a) and Stokes-anti-Stokes coincidence probability (b) as a function of the mode number $m$.

The probability of generating an atom–photon (photon–photon) entanglement pair corresponds to the total Stokes detection (Stokes–anti-Stokes coincidence) probability $P_S^{(m)} = \sum_{i=1}^{m} P_S^{i\text{-th}}$ ($P_{S,T}^{(m)} = \sum_{i=1}^{m} P_{S,T}^{i\text{-th}}$). The blue squares (circles) dots in Fig. 4a (Fig. 4b) are the measured values of $P_S^{(m)}$ ($P_{S,T}^{(m)}$) as a function of $m$ and show that the MQI gives rise to a three-fold increase in the atom–photon (photon–photon) entanglement-generation probability compared with single-mode QIs. Considering the imperfect OSN efficiency $\eta_{sw} \approx 0.8$, which remains fixed regardless of $m$ [55], the MQI increases the photon–photon entanglement-generation probability by a factor of $m \times \eta_{sw} = 2.4$ compared with the single-mode QI without OSN.

We have demonstrated a three-mode MQI that preserves SWPE for 1

ms. This lifetime is 20 times longer than the best results among the multimode SWPE QIs reported to date. If the three-mode MQIs instead of single-mode MQIs are used as nodes of an elementary link, the probability of entanglement generation in the link will be increased 3-fold. To apply the present MQI in QR applications, its performances needs to be further improved. Millisecond lifetimes are mainly limited by motional dephasing (see Supplementary Material [68]) but can be prolonged to 0.2 s by trapping the atoms in an optical lattice [25, 26]. The multimode number can be increased by extending the apertures of the optical devices. The multimode capacity may be extended further using the multiplexing schemes with two or more degrees of freedom [76, 77], e.g., combining a temporal multiplexing scheme [65] with the present spatial approach. Considering an MQI that stores 65 spatial and 10 temporal spin-wave qubits, the total number of memory qubits reaches $N_m = 650$. To minimize transmission losses in the fibers, the Stokes photons at Rb transitions may be converted into photons in the telecommunications band [48, 78–80]. The lower retrieval efficiency (15%) can be increased using high optical-depth cold atoms [81] or coupling the atoms with an optical cavity [23, 38]. We remark that the recent experiment demonstrates probabilistic entanglement generation in an elementary link over 22-*km* field-deployed fiber [48], where the link uses single-mode SWPE QIs as nodes. The deterministic entanglement generation in this link [82] requires very long storage time (~150 $s$) [48], which lead to very low rate ($\sim 6.7\times 10^{-3}$ Hz). More importantly, the required storage times are far beyond the art-of-state lifetime ($T_{as} = 0.22$ $s$) [26] and don't allow one to implement the deterministic generation with current technology. To overcome these problems, one may use the MQIs storing $N_m = 650$ qubits instead of the single-mode QIs, which reduces the required storage time to $150 s / N_m = 0.23$ $s$ [56] and then increases the rate to 4.3 Hz. With the required storage time reaching to the order $T_{as}$, one may implement the deterministic generation.

Our present experiment shows a promising way to store all of the multiplexed memory qubits as MFI spin waves, thereby allowing realization of an entanglement QI capable of storing a large number of long-lived memory qubits in cold atoms, thus benefitting QR-based long-distance quantum communications.

*Corresponding author: wanghai@sxu.edu.cn

We acknowledge funding support from Key Project of the Ministry of Science and Technology of China (Grant No. 2016YFA0301402), the National Natural Science Foundation of China (Grants No. 11475109, 11974228 ), and Shanxi "1331 Project". We also thank Richard Haase, Ph.D, from Liwen Bianji, Edanz Group China (www.liwenbianji.cn/ac),

for editing the English text a draft of this manuscript.

References:

[1] H. J. Briegel, W. Dür, J. I. Cirac, and P. Zoller, Physical Review Letters **81**, 5932 (1998).

[2] N. Sangouard, C. Simon, H. de Riedmatten, and N. Gisin, Reviews of Modern Physics **83**, 33 (2011).

[3] C. Simon, Nature Photonics **11**, 678 (2017).

[4] F. Bussieres, N. Sangouard, M. Afzelius, H. de Riedmatten, C. Simon, and W. Tittel, Journal of Modern Optics **60**, 1519 (2013).

[5] H. J. Kimble, Nature **453**, 1023 (2008).

[6] S. Wehner, D. Elkouss, and R. Hanson, Science **362**, eaam9288 (2018).

[7] L. M. Duan, M. D. Lukin, J. I. Cirac, and P. Zoller, Nature **414**, 413 (2001).

[8] A. I. Lvovsky, B. C. Sanders, and W. Tittel, Nature Photonics **3**, 706 (2009).

[9] J. Volz, M. Weber, D. Schlenk, W. Rosenfeld, J. Vrana, K. Saucke, C. Kurtsiefer, and H. Weinfurter, Phys Rev Lett **96**, 030404 (2006).

[10] A. Reiserer and G. Rempe, Reviews of Modern Physics **87**, 1379 (2015).

[11] B. B. Blinov, D. L. Moehring, L. Duan, and C. Monroe, Nature **428**, 153 (2004).

[12] L. M. Duan and C. Monroe, Reviews of Modern Physics **82**, 1209 (2010).

[13] B. Hensen *et al.*, Nature **526**, 682 (2015).

[14] A. Delteil, Z. Sun, W.-b. Gao, E. Togan, S. Faelt, and A. Imamoğlu, Nature Physics **12**, 218 (2015).

[15] A. Kuzmich, W. P. Bowen, A. D. Boozer, A. Boca, C. W. Chou, L. M. Duan, and

H. J. Kimble, Nature **423**, 731 (2003).

[16] C. W. Chou, H. de Riedmatten, D. Felinto, S. V. Polyakov, S. J. van Enk, and H. J. Kimble, Nature **438**, 828 (2005).

[17] M. D. Eisaman, A. Andre, F. Massou, M. Fleischhauer, A. S. Zibrov, and M. D. Lukin, Nature **438**, 837 (2005).

[18] J. Laurat, H. de Riedmatten, D. Felinto, C. W. Chou, E. W. Schomburg, and H. J. Kimble, Opt Express **14**, 6912 (2006).

[19] J. Simon, H. Tanji, J. K. Thompson, and V. Vuletic, Phys Rev Lett **98**, 183601 (2007).

[20] D. Felinto, C. W. Chou, H. de Riedmatten, S. V. Polyakov, and H. J. Kimble, Physical Review A **72**, 053809 (2005).

[21] J. Laurat, K. S. Choi, H. Deng, C. W. Chou, and H. J. Kimble, Phys Rev Lett **99**, 180504 (2007).

[22] B. Zhao *et al.*, Nature Physics **5**, 95 (2009).

[23] X. H. Bao *et al.*, Nature Physics **8**, 517 (2012).

[24] R. Zhao, Y. O. Dudin, S. D. Jenkins, C. J. Campbell, D. N. Matsukevich, T. A. B. Kennedy, and A. Kuzmich, Nature Physics **5**, 100 (2009).

[25] A. G. Radnaev, Y. O. Dudin, R. Zhao, H. H. Jen, S. D. Jenkins, A. Kuzmich, and T. A. B. Kennedy, Nature Physics **6**, 894 (2010).

[26] S.-J. Yang, X.-J. Wang, X.-H. Bao, and J.-W. Pan, Nature Photonics **10**, 381 (2016).

[27] X. L. Pang *et al.*, Sci Adv **6**, eaax1425 (2020).

[28] M. Zugenmaier, K. B. Dideriksen, A. S. Sørensen, B. Albrecht, and E. S. Polzik, Communications Physics **1**, 76 (2018).

[29] D. N. Matsukevich and A. Kuzmich, Science **306**, 663 (2004).

[30] H. de Riedmatten, J. Laurat, C. W. Chou, E. W. Schomburg, D. Felinto, and H. J. Kimble, Phys Rev Lett **97**, 113603 (2006).

[31] S. Chen, Y. A. Chen, B. Zhao, Z. S. Yuan, J. Schmiedmayer, and J. W. Pan, Phys Rev Lett **99**, 180505 (2007).

[32] K. Kutluer, E. Distante, B. Casabone, S. Duranti, M. Mazzera, and H. de Riedmatten, Phys Rev Lett **123**, 030501 (2019).

[33] B. Zhao, Z. B. Chen, Y. A. Chen, J. Schmiedmayer, and J. W. Pan, Phys Rev Lett **98**, 240502 (2007).

[34] Z. B. Chen, B. Zhao, Y. A. Chen, J. Schmiedmayer, and J. W. Pan, Physical Review A **76**, 022329 (2007).

[35] Z. S. Yuan, Y. A. Chen, B. Zhao, S. Chen, J. Schmiedmayer, and J. W. Pan, Nature **454**, 1098 (2008).

[36] D. N. Matsukevich, T. Chaneliere, M. Bhattacharya, S. Y. Lan, S. D. Jenkins, T. A. Kennedy, and A. Kuzmich, Phys Rev Lett **95**, 040405 (2005).

[37] Y. O. Dudin, A. G. Radnaev, R. Zhao, J. Z. Blumoff, T. A. Kennedy, and A. Kuzmich, Phys Rev Lett **105**, 260502 (2010).

[38] S. J. Yang, X. J. Wang, J. Li, J. Rui, X. H. Bao, and J. W. Pan, Phys Rev Lett **114**, 210501 (2015).

[39] D. S. Ding, W. Zhang, Z. Y. Zhou, S. Shi, B. S. Shi, and G. C. Guo, Nature

Photonics **9**, 332 (2015).

[40] Y. Wu *et al.*, Physical Review A **93**, 052327 (2016).

[41] P. Farrera, G. Heinze, and H. de Riedmatten, Phys Rev Lett **120**, 100501 (2018).

[42] E. Saglamyurek *et al.*, Nature **469**, 512 (2011).

[43] C. Clausen, I. Usmani, F. Bussieres, N. Sangouard, M. Afzelius, H. de Riedmatten, and N. Gisin, Nature **469**, 508 (2011).

[44] E. Saglamyurek, J. Jin, V. B. Verma, M. D. Shaw, F. Marsili, S. W. Nam, D. Oblak, and W. Tittel, Nature Photonics **9**, 83 (2015).

[45] Y. A. Chen, S. Chen, Z. S. Yuan, B. Zhao, C. S. Chuu, J. Schmiedmayer, and J. W. Pan, Nature Physics **4**, 103 (2008).

[46] F. Bussières *et al.*, Nature Photonics **8**, 775 (2014).

[47] X. H. Bao, X. F. Xu, C. M. Li, Z. S. Yuan, C. Y. Lu, and J. W. Pan, Proc Natl Acad Sci U S A **109**, 20347 (2012).

[48] Y. Yu *et al.*, Nature **578**, 240 (2020).

[49] M. Razavi, M. Piani, and N. Lutkenhaus, Physical Review A **80**, 032301 (2009).

[50] L. Jiang, J. M. Taylor, and M. D. Lukin, Physical Review A **76**, 012301 (2007).

[51] N. Sangouard, C. Simon, B. Zhao, Y. A. Chen, H. de Riedmatten, J. W. Pan, and N. Gisin, Physical Review A **77**, 062301 (2008).

[52] C. Simon, H. de Riedmatten, M. Afzelius, N. Sangouard, H. Zbinden, and N. Gisin, Phys Rev Lett **98**, 190503 (2007).

[53] O. A. Collins, S. D. Jenkins, A. Kuzmich, and T. A. Kennedy, Phys Rev Lett **98**,

060502 (2007).

[54] C. Simon, H. de Riedmatten, and M. Afzelius, Physical Review A **82**, 010304(R) (2010).

[55] N. Sinclair *et al.*, Phys Rev Lett **113**, 053603 (2014).

[56] L. Tian, Z. Xu, L. Chen, W. Ge, H. Yuan, Y. Wen, S. Wang, S. Li, and H. Wang, Phys Rev Lett **119**, 130505 (2017).

[57] C. Laplane, P. Jobez, J. Etesse, N. Gisin, and M. Afzelius, Phys Rev Lett **118**, 210501 (2017).

[58] K. Kutluer, M. Mazzera, and H. de Riedmatten, Phys Rev Lett **118**, 210502 (2017).

[59] A. Seri, A. Lenhard, D. Rieländer, M. Gündoğan, P. M. Ledingham, M. Mazzera, and H. de Riedmatten, Physical Review X **7**, 021028 (2017).

[60] B. Albrecht, P. Farrera, G. Heinze, M. Cristiani, and H. de Riedmatten, Phys Rev Lett **115**, 160501 (2015).

[61] L. Heller, P. Farrera, G. Heinze, and H. de Riedmatten, Physical Review Letters **124**, 210504 (2020).

[62] M. Parniak, M. Dabrowski, M. Mazelanik, A. Leszczynski, M. Lipka, and W. Wasilewski, Nat Commun **8**, 2140 (2017).

[63] S. Y. Lan, A. G. Radnaev, O. A. Collins, D. N. Matsukevich, T. A. Kennedy, and A. Kuzmich, Opt Express **17**, 13639 (2009).

[64] Y. F. Pu, N. Jiang, W. Chang, H. X. Yang, C. Li, and L. M. Duan, Nat Commun **8**, 15359 (2017).

[65] W. Chang, C. Li, Y. K. Wu, N. Jiang, S. Zhang, Y. F. Pu, X. Y. Chang, and L. M. Duan, Physical Review X **9**, 041033 (2019).

[66] Y. Wen, P. Zhou, Z. Xu, L. Yuan, H. Zhang, S. Wang, L. Tian, S. Li, and H. Wang, Physical Review A **100**, 012342 (2019).

[67] K. R. Ferguson, S. E. Beavan, J. J. Longdell, and M. J. Sellars, Phys Rev Lett **117**, 020501 (2016).

[68] see Supplementary Material.

[69] P. Vernaz-Gris, K. Huang, M. Cao, A. S. Sheremet, and J. Laurat, Nat Commun **9**, 363 (2018).

[70] Y. Wang, J. Li, S. Zhang, K. Su, Y. Zhou, K. Liao, S. Du, H. Yan, and S.-L. Zhu, Nature Photonics **13**, 346 (2019).

[71] D. G. England *et al.*, J Phys B-at Mol Opt **45**, 124008 (2012).

[72] Y. W. Cho and Y. H. Kim, Opt Express **18**, 25786 (2010).

[73] Y. W. Cho *et al.*, Optica **3**, 100 (2016).

[74] S. Wang, Z. Xu, D. Wang, Y. Wen, M. Wang, P. Zhou, L. Yuan, S. Li, and H. Wang, Opt Express **27**, 27409 (2019).

[75] Z. Xu *et al.*, Phys Rev Lett **111**, 240503 (2013).

[76] T. S. Yang *et al.*, Nat Commun **9**, 3407 (2018).

[77] A. Seri, D. Lago-Rivera, A. Lenhard, G. Corrielli, R. Osellame, M. Mazzera, and H. de Riedmatten, Phys Rev Lett **123**, 080502 (2019).

[78] B. Albrecht, P. Farrera, X. Fernandez-Gonzalvo, M. Cristiani, and H. de Riedmatten, Nat Commun **5**, 3376 (2014).

[79] R. Ikuta *et al.*, Nat Commun **9**, 1997 (2018).

[80] T. van Leent, M. Bock, R. Garthoff, K. Redeker, W. Zhang, T. Bauer, W. Rosenfeld, C. Becher, and H. Weinfurter, Phys Rev Lett **124**, 010510 (2020).

[81] S. Zhang, J. F. Chen, C. Liu, S. Zhou, M. M. Loy, G. K. Wong, and S. Du, Rev Sci Instrum **83**, 073102 (2012).

[82] P. C. Humphreys, N. Kalb, J. P. J. Morits, R. N. Schouten, R. F. L. Vermeulen, D. J. Twitchen, M. Markham, and R. Hanson, Nature **558**, 268 (2018).

## Supplementary Material

**Entanglement establishment between two remote nodes in a "heralded" fashion.**

As explained in the main text, a DLCZ-like elementary link comprises two nodes, each being formed by a QI. One essential requirement for the QR protocols is to establish entanglement between the two nodes [2, 47]. The basic process to establish the entanglement in a "heralded" fashion includes: (1) sending the photons from each QI to the center station between the two nodes to perform a Bell-state measurement (BSM); (2) resending the information about the BSM result to the memories in the QIs; and (3) depending on the result of the BSM received, knowing whether at each node the entanglement may be established between the two memories is successful and then determining the next step. If it is successful, the memory will continuously preserve the entanglement; if not, the memory has to be emptied and the next attempt made. The time interval required for one attempt is $\Delta t = L_0/c$, where $L_0$ denotes the separated distance of the two nodes in the elementary link, and $c = 2 \times 10^8 m/s$ the speed of light in the optical fibers. This attempt requires that the memory stores spin waves for the same duration $L_0/c$, which in turn requires that the quantum memory has a lifetime of over $L_0/c$. In this way, if two DLCZ-like QMs with lifetimes of $\tau_0 \sim 50\,\mu s$ are placed at two separated nodes by a distance $L_0 \geq 10\,km$, one cannot attempt to generate entanglement between them in the "heralded" manner because $\tau_0 \leq L_0/c$.

If MQIs that generate $m$-mode entanglement between a spin wave and a photon are used as nodes in an elementary link, the entanglement-generation probability in the elementary link increases $m$-fold compared with single-mode links [56]. We emphasize, however, that this increase can be achieved only when the entanglement in the multiplexed link is established in the "heralded" manner. The reason for this is that the multimode memory at each node needs to know that successful BSMs have been achieved and in which photonic modes so that it can retrieve the excitations in the corresponding spin-wave modes and send them to a common channel via feed-forward-controlled read outs [56]. According to the above discussions, one cannot realize a probability increase in the entanglement generation between two remote multimode QMs linked by more than 10-$km$ fibers because of the short storage lifetimes (~50 $\mu s$).

Our present three-mode QM has a lifetime of $\tau_0 \sim 1ms$ for preserving entanglement. If one want to establish entanglement between the two QMs in the "heralded" manner, the fiber distance used for connecting them can reach up to the order of $L_0 \sim \tau_0 c = 200km$.

**Experimental details.** The experiment is carried out in a cyclic fashion. In each experimental circle, the durations for the preparation of cold atoms and the experiment run of the SWPE generation are 42 *ms* and 8 *ms*, respectively, corresponding to a 20-Hz repetition rate. During the preparation stage, more than $10^8$ $^{87}$Rb atoms are trapped in a two-dimension magneto-optical trap (MOT) for 41.5 *ms* and further cooled by a Sisyphus cooling for 0.5 *ms*. The cloud of the cold atoms has a size of ~5×2×2 $mm^3$, a temperature of ~100 $\mu K$ and an optical density of about 14. At the end of this preparation stage, a bias magnetic field of $B_0$=4 *G* is applied along *z*-axis (see Fig.1a) and the atoms are prepared into the initial level $|5^2 S_{1/2}, F=1, m=-1\rangle$ via optical pumping[74]. After the preparation stage, the experimental run containing many SWPE generation trials start. In the center of the atoms, the diameters of the write and read light beams are all ~1.1 *mm*, but the powers of them are 100 $\mu W$ and ~1 *mW*, respectively. The read light field is on resonance with the transition $|b\rangle \to |e_1\rangle$. The durations of the write and read pulses are all 70 *ns*. At the right site in the Fig.1a, the BD2 perfectly combines the $A_1^{(i)}$ (*H*-polarization) and $A_2^{(i)}$ (*V*-polarization) modes into a single light beam propagating in OC$_i$ and then the light beam is effectively coupled into the single-mode fiber $SMF_S^{(i)}$. The measured coupling efficiencies for the modes $\{A_1^{(1)}, A_2^{(1)}\}$, $\{A_1^{(2)}, A_2^{(2)}\}$ and $\{A_1^{(3)}, A_2^{(3)}\}$ are {70.5%, 71.0%} {70.6%, 71.5%} and {70.8%, 70%}, respectively. For blocking the write

(read) beam in the Stokes and anti-Stokes channels, we place an optical-spectrum-filter set (OSFS) before each polarization beam splitter (PBS). Each OSFS is consisted of four F-P etalons, which can attenuate the write (read) beam by a factor of $2.7\times10^{-9}$ ($3.7\times10^{-9}$) and transmit the Stokes ((anti-Stokes) fields with a transmission of ~65%. Also, in the Stokes (anti-Stokes) detection channel, the spatial separation of the Stokes (anti-Stokes fields) from the strong write (read) beam provides an attenuation of ~$10^{-4}$ for the write (read) beam.

In the spontaneously Raman emission induced by the write pulse, if the Stokes photon $S_1^{(i)}$ ($S_2^{(i)}$) is $\sigma^+$-polarization, it will be transformed into $V$ ($H$) –polarized photon by the λ/4 wave-plate QW$_{1R}$ (QW$_{2R}$) (see Fig.1a) and then removed from the OC$_i$ by BD2.

All error bars in the experimental data represent $\pm 1$ standard deviation, which are estimated from Poissonian detection statistic using Monte Carlo simulation.

**Emission directions of the retrieved photons.** Thanks to collective interference of the atoms, the retrieved photon $T_1^{(i)}$ ($T_2^{(i)}$) is emitted into a well-defined spatial mode given by the phase matching condition, $k_{T_1}^{(i)} = k_{M_1}^{(i)} + k_R = k_W - k_{S_1}^{(i)} + k_R$ ($k_{T_2}^{(i)} = k_{M_2}^{(i)} + k_R = k_W - k_{S_2}^{(i)} + k_R$), where, $k_R$ is the wave-vector of the read beam. Since the write and read light beams counter-propagate through the atoms, we have $k_R \approx -k_w$ and then have $k_{T_1}^{(i)} \approx -k_{S_1}^{(i)}$ ($k_{T_2}^{(i)} \approx -k_{S_2}^{(i)}$).

**Phase compensators (PCs)**. When *H*- and *V*- polarization light fields respectively propagates in the two paired arms in the BDs, the refractive index difference between the two arms will lead to a phase shift between the two light fields. We use PCs to overcome the problems. For example, for eliminating the phase shift due to the BDs in the *i*-th channel ($OC_i$), we place the phase compensator $PC_i$ between BD1 and the *i*-th single-mode fiber at the left site. Each phase compensators is a combination of λ/4, λ/2 and λ/4 wave-plates [56]. By rotating the λ/2 wave-plate in the $PC_i$, we can compensate the phase shift due to the BDs ($\varphi_i + \psi_i$) to zero. We also insert a PC before each PBS in the Fig.1a to eliminate phase shifts due to the optical elements such as single-mode fibers and AOMs.

**The increase in the multimode number**

To increase the multimode number in the present experimental setup, we have to extend the apertures of the optical devices including BDs, BTDs and the lenses. When the apertures of the optical elements, for example, are all extended to $48 \times 48$ $mm^2$, a $13 \times 10 = 130$ light-beam array which corresponds to the storage of 65 MFI spin-wave qubits will be implemented.

# The configuration of the PI and cross sections of the array at different sites.

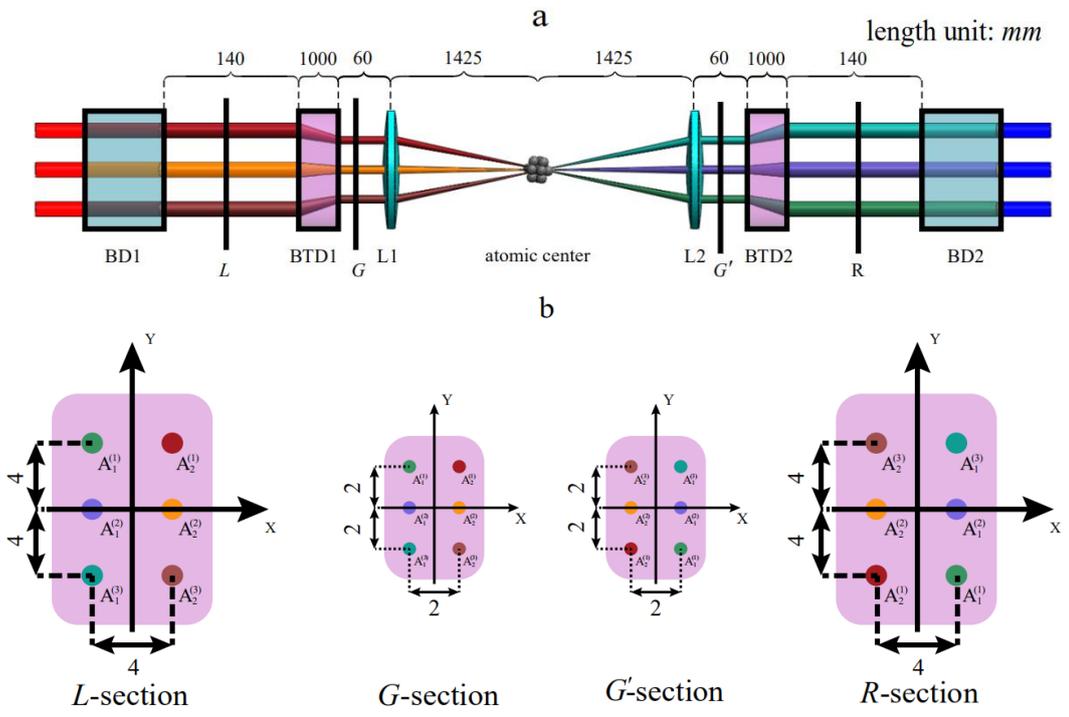

Supplementary Figure 1. The configuration of the PI (a) and $L, G, G'$ and $R$ -section of the array

# Beam transformation devices (BTDs).

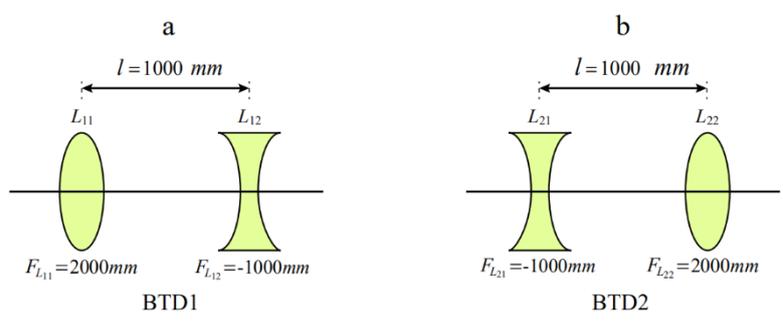

Supplementary Figure 2. The configurations of the BTD1 (a) and BTD2 (b).

We use two BTDs (BTD1 and BTD2) to transform the beam array in the presented experiment. The BTD1 (BTD2) is formed by a convex and a concave lens, which are denoted by $L_{11}$ ($L_{22}$) and $L_{12}$ ($L_{21}$), respectively.

The configuration of the BTD1 (BTD2) is shown in FigS.2a (FigS.2b), where the convex (concave) lens $L_{11}$ ($L_{12}$) is placed at the left side. The focus length of the lens $L_{11}$ ($L_{22}$) and $L_{12}$ ($L_{21}$) are selected to be as $F_{L_{11}} = f_L$ ($F_{L_{22}} = f_L$) and $F_{L_{12}} = -f_L/F$ ($F_{L_{21}} = -f_L/F$), with $f_L > 0$. The separation between $L_{11}$ ($L_{22}$) and $L_{12}$ ($L_{21}$) is equal to $(1-1/F)f_L$. The ABCD matrix for the BTD1 and BTD2 can be calculated by:

$$\begin{pmatrix} 1 & 0 \\ F/f_L & 1 \end{pmatrix} \begin{pmatrix} 1 & (1-1/F)f_L \\ 0 & 1 \end{pmatrix} \begin{pmatrix} 1 & 0 \\ -1/f_L & 1 \end{pmatrix} = \begin{pmatrix} 1/F & (1-1/F)f_L \\ 0 & F \end{pmatrix},$$

and

$$\begin{pmatrix} 1 & 0 \\ -1/f_L & 1 \end{pmatrix} \begin{pmatrix} 1 & (1-1/F)f_L \\ 0 & 1 \end{pmatrix} \begin{pmatrix} 1 & 0 \\ F/f_L & 1 \end{pmatrix} = \begin{pmatrix} F & (1-1/F)f_L \\ 0 & 1/F \end{pmatrix}$$

Thus, when a beam array goes through the BTD1 (BTD2) from the left to right side, it can be shrunk (expand) by a factor $F$. In our current experiment, we use the lenses at hand, whose focus lengths are $F_{L_{11}} = 2000$ mm, $F_{L_{22}} = 2000$ mm, $F_{L_{12}} = -1000$ mm and $F_{L_{21}} = -1000$ mm, which corresponds to the factor $F=2$. In the next works, one may increase (decrease) the factor $F$ ($1/F$) by increasing (decrease) the lens ratio $|F_{L_{11}}/F_{L_{12}}|$ ($|F_{L_{22}}/F_{L_{21}}|$).

**The angles of the six modes**

The angles $\vartheta_\alpha^{(i)}$ of the six modes $A_\alpha^{(i)}$ ($\alpha=1, 2$; $i=1$ to $3$) relative to the write beam which propagates along the z-axis. The G-section of Supplementary Figure 1 shows the cross section on the lens L1, the distance from the center of the mode $A_\alpha^{(i)}$ to the $x=y=0$ point is

$d_\alpha^{(i)} = \sqrt{\left(x_\alpha^{(i)}\right)^2 + \left(y_\alpha^{(i)}\right)^2} = \left(\sqrt{4(i-2)^2 + 1}\right) B_f / 2$, where $B_f$ is the beam separation of the cross section. So, the angles $\vartheta_\alpha^{(i)}$ can be calculated by $\vartheta_\alpha^{(i)} = \operatorname{tg}^{-1}\left(d_\alpha^{(i)} / z_0\right)$, where $z_0$ is the length from the center of L1 to that of the atoms, which is equal to the focus length $f$ of the L1. Since $d_\alpha^{(i)} \ll f$, we have $\vartheta_\alpha^{(i)} \approx \left(d_\alpha^{(i)} / z_0\right) = \left(\sqrt{4(i-2)^2 + 1}\right) B_f / 2f$.

**The lifetimes limited by the atomic motion.**

For the storage of a single-mode spin wave, it has been pointed out the decoerhence due to the atomic motions give a limit to the storage lifetimes, which may be described by $\mathrm{T} \approx \Lambda / \bar{v}$ [2, 73], where, $\bar{v} = \sqrt{k_B T / m}$ is the average atomic speed, $\Lambda \approx 2\pi / k_W \vartheta$ is the wavelengths of the spin wave, $\vartheta$ is the angle of the Stokes photon relative to the write beam (z-axis) and $k_W$ is the wave-vector of the write beam. Returning to the present multimode storage, the wavelengths of the spin-wave $M_\alpha^{(i)}$ may be written as: $\Lambda_\alpha^{(i)} \approx 2\pi / k_W \vartheta_\alpha^{(i)}$, and the storage lifetimes limited to the motion-induced-decoherence are $\mathrm{T}_\alpha^{(i)} \approx \Lambda_\alpha^{(i)} / \bar{v}$ for $M_\alpha^{(i)}$ modes. For the present experiment, $T \simeq 100\ \mu K$ and { $\vartheta_{A_1}^{(1)} = \vartheta_{A_2}^{(1)} \approx 0.09°$, $\vartheta_{A_1}^{(2)} = \vartheta_{A_2}^{(2)} \approx 0.04°$, $\vartheta_{A_1}^{(3)} = \vartheta_{A_2}^{(3)} \approx 0.09°$ }, we obtain { $\mathrm{T}_1^{(1)} = \mathrm{T}_2^{(1)} \simeq 840\ \mu s$, $\mathrm{T}_1^{(2)} = \mathrm{T}_2^{(2)} \simeq 1850\ \mu s$, $\mathrm{T}_1^{(3)} = \mathrm{T}_2^{(3)} \simeq 840\ \mu s$ }.

**The retrieval efficiencies of the individual spin-wave qubits as function of storage time.**

We also measure the retrieval efficiencies of the 1$^{th}$, 2$^{th}$ and 3$^{th}$ spin-wave (SW) qubits as a function of $t$, which are shown as circle, diamond and square dots in Fig.S3. The dash and solid curves are the fitting of the measured retrieval efficiencies of the 1$^{th}$ (3$^{th}$) and 2$^{th}$ spin-wave qubits based on $\gamma^{(1-th)} = \gamma_0 e^{-t/\tau_0^{(1-th)}}$ ( $\gamma^{(3-th)} = \gamma_0 e^{-t/\tau_0^{(3-th)}}$ ) and $\gamma^{(2-th)} = \gamma_0 e^{-t/\tau_0^{(2-th)}}$, which yield the 1/e storage lifetimes of $\tau_0^{(1-th)} = \tau_0^{(3-th)} = 730\ \mu s$, and $\tau_0^{(2-th)} = 1170\ \mu s$, respectively. The average lifetimes over the three spin-wave qubits are $\bar{\tau}_0 = \left(\tau_0^{(1-th)} + \tau_0^{(2-th)} + \tau_0^{(3-th)}\right)/3 \approx 876\ \mu s$, which is agreement with results in the Fig.2 of the main text.

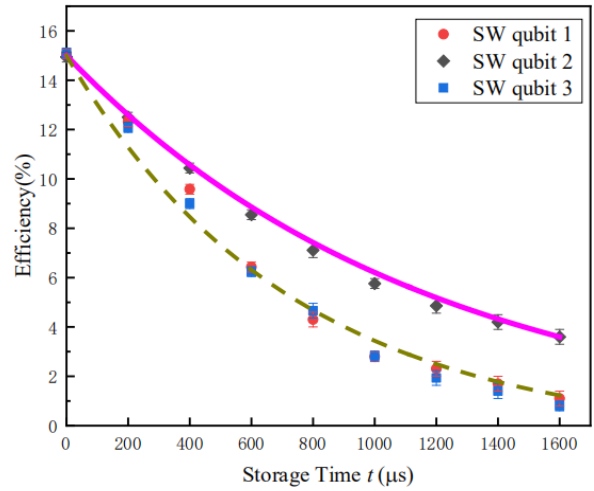

Supplementary Figure 3. Retrieval efficiencies of the three spin-wave (SW) qubits as a function of storage time $t$.

**The analysis on the main limits to the QM lifetimes in the presented experiment.**

As explained in the main text, the lifetime of a spin wave are limited by two main factors, one is atomic motions and the other is inhomogeneous broadening of the spin transitions. Consider the two factors, the total lifetime for $i$-th SW can be described by $\tau_0^{(i)} = \mathrm{T}^{(i)} \tau_m / (\mathrm{T}^{(i)} + \tau_m)$ [74], where, $\mathrm{T}^{(i)}$ is the lifetime limited by the decoherence due to atomic motions, which is described and given by the above, $\tau_m$ is the lifetime limited by the decoherence due to the inhomogeneous broadening of the spin transitions. Such inhomogeneous broadening is caused by the spatial gradient $B'$ of the magnetic field $B$, which equals to the Zeeman frequency splitting across the ensemble. For the spin-wave storage based on Zeeman coherence $|a, m_a\rangle \leftrightarrow |b, m_b\rangle$, the Zeeman splitting can be described by the parameter $K = (|g_a(m_a + m_b) + \delta g m_b| \mu_B l B') / h$ [29], where, $g_a \approx 0.5018$, $\delta g = g_a + g_b = -0.002$, $h$ is Planck's Constant, $\mu_B$ is Bohr Magneton. So, the lifetime $\tau_m$ can be evaluated according to $\tau_m \propto 1/K = h / (|g_a(m_a + m_b) + \delta g m_b| \mu_B l B')$. For the present experiment, $B' \approx 22 \, mG/cm$, $l = 5 \, mm$, we obtain $\tau_m \approx 32 \, ms$, which is much longer than $\mathrm{T}_\alpha^{(i)}$. Thus, the lifetime $\tau_0$ ($\bar{\tau}_0$) in the present experiment is mainly limited by the decoherence due to the atomic motions.

**The reconstructed density matrixes of the two-photon entangled states.**

For reconstructing the density matrix $\rho_r^{(m)}$, we measure the Stokes-anti-Stokes coincidence count rates $C_{D_1, T_1}^{(m)}(X, Y)$, $C_{D_1, T_2}^{(m)}(X, Y')$,

$C_{D_2,T_1}^{(m)}(X',Y)$, $C_{D_2,T_2}^{(m)}(X',Y')$, where, for example,

$$C_{D_1,T_1}^{(m)}(X,Y) = \sum_{i=1}^{m} C_{D_1,T_1}^{(i\text{-th})}(X,Y) \quad (C_{D_2,T_2}^{(m)}(X',Y') = \sum_{i=1}^{m} C_{D_2,T_2}^{(i\text{-th})}(X',Y')),$$

$C_{D_1,T_1}^{(i\text{-th})}(X,Y)$ ($C_{D_2,T_2}^{(i\text{-th})}(X',Y')$) is the coincidence count rate between the detectors $D_{S_1}^{(i)}$ ($D_{S_2}^{(i)}$) that measures the X-polarized (X'-polarized) Stokes photon and $D_{T_1}$ ($D_{T_2}$) that measures Y-polarized (Y'-polarized) anti-Stokes photon, $X$ ($Y$) = $H, D$ or $\sigma^+$ ($X'$ ($Y'$) = $V, A$ or $\sigma^-$), H and V represent horizontal and vertical polarizations, D and A represents $+45^0$ and $-45^0$ linear polarizations, $\sigma^+$ and $\sigma^-$ represents right and left circular polarizations, respectively. The detections of $X(Y) = H, D$ -polarized photons at $D_{S_1}^{(i)}$ ($D_{T_1}$), which corresponds to the simultaneous detections of $X'(Y') = V, A$ -polarized photons at $D_{S_2}^{(i)}$ ($D_{T_2}$), are accomplished by setting the polarization angles of the λ/2 wave-plates before the PBSs. When performing the detections of the $X(Y) = \sigma^+$ -polarized photon at $D_{S_1}^{(i)}$ ($D_{T_1}$), which corresponds to the simultaneous detections of $X'(Y') = \sigma^-$ at the detectors $D_{S_2}^{(i)}$ ($D_{T_2}$), we use λ/4 wave-plates to replace the λ/2 wave plates. Then, by setting the polarization angles of the λ/4 wave-plates, we achieve the detections of $\sigma^+$ and $\sigma^-$-polarized photons. Based on the measured Stokes-anti-Stokes coincidence rates, we reconstruct the density matrix $\rho_r^{(m)}$ and plot it in Supplementary Figure 4, which yields $F^{(m=3)} = 90.4 \pm 1.6\%$.

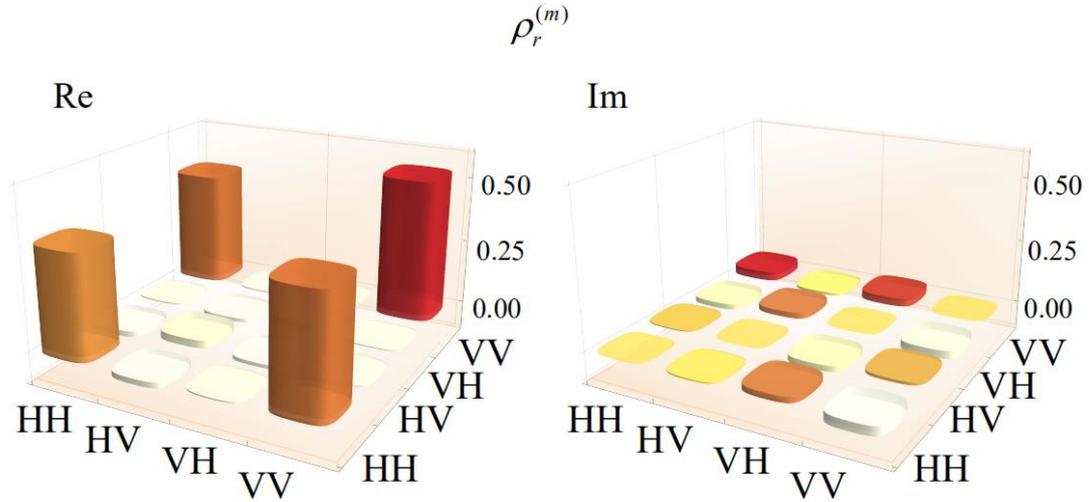

Supplementary Figure 4. Real and imaginary parts of the reconstructed density matrix $\rho_r^{(m)}$

We also measure the individual Stokes-anti-Stokes coincidence count rates $C_{D_1,T_1}^{(i\text{-th})}(X,Y)$, $C_{D_1,T_2}^{(i\text{-th})}(X,Y')$, $C_{D_2,T_1}^{(i\text{-th})}(X',Y)$, $C_{D_2,T_2}^{(i\text{-th})}(X',Y')$ for the different polarization cases and then give the reconstructed density matrices $\rho_r^{i-\text{th}}$ of the two-photon entangled states $\Phi_{pp}^{(i)}$ ($i=1,2,3$), respectively. Supplementary Figure 5 plots the real and imaginary parts of the density matrices $\rho_r^{1\text{-th}}$, $\rho_r^{2\text{-th}}$ and $\rho_r^{3\text{-th}}$, which yield $F^{(1\text{-th})}=88.6\pm1.13\%$, $F^{(2\text{-th})}=92.0\pm1.5\%$ and $F^{(3\text{-th})}=88.4\pm0.85\%$, respectively.

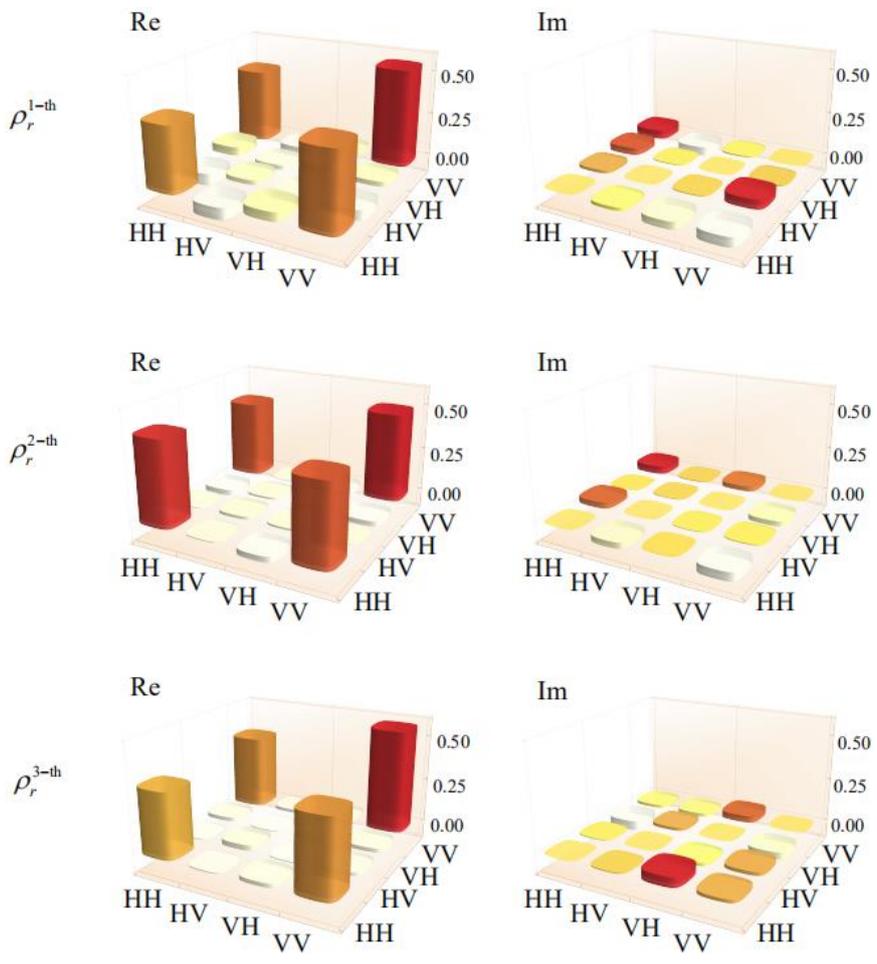

Supplementary Figure 5. Real and imaginary parts of the reconstructed density matrices $\rho_r^{1\text{-th}}$, $\rho_r^{2\text{-th}}$ and $\rho_r^{3\text{-th}}$, respectively.